\documentclass[a4paper,superscriptaddress,preprintnumbers,floatfix,letter]{elsarticle}
\usepackage{bm}
\usepackage{amsmath}
\usepackage{amssymb}
\usepackage{amsthm}
\theoremstyle{definition}
\newtheorem{theorem}{Theorem}
\newtheorem*{theorem*}{Theorem}

\newtheorem*{definition*}{Definition}

\begin{document}
\title{Noise-induced degeneration in online learning}
\author{Yuzuru Sato}
\address{RIES / Department of Mathematics, Hokkaido University, Kita 20 Nishi 10, Kita-ku, Sapporo 001-0020, Japan}
\address{London Mathematical Laboratory, 8 Margravine Gardens, London W68RH, UK}
\author{Daiji Tsutsui}
\address{Department of Mathematics, Osaka University, Toyonaka, Osaka 560-0043, JAPAN}
\author{Akio Fujiwara}
\address{Department of Mathematics, Osaka University, Toyonaka, Osaka 560-0043, JAPAN}

\begin{abstract}
    In order to elucidate the plateau phenomena caused by vanishing gradient, we herein analyse stability of stochastic gradient descent near degenerated subspaces in a multi-layer perceptron. In stochastic gradient descent for Fukumizu-Amari model, which is the minimal multi-layer perceptron showing non-trivial plateau phenomena, we show that (1) attracting regions exist in multiply degenerated subspaces, (2) a strong plateau phenomenon emerges as a noise-induced synchronisation, which is not observed in deterministic gradient descent, (3) an optimal fluctuation exists to minimise the escape time from the degenerated subspace. The noise-induced degeneration observed herein is expected to be found in a broad class of machine learning via neural networks. 
\end{abstract}
\maketitle

\section{Introduction}

  The least square learning of neural networks is a typical framework in machine learning. The gradient descent is the simplest optimisation algorithm represented by gradient dynamics in a potential. When the input data is finite, gradient descent dynamics fluctuates due to the finite size effects, and is called stochastic gradient descent. In this paper, we study stability of stochastic gradient descent dynamics from the viewpoint of dynamical systems theory. 
  
  Learning is characterised as non-autonomous dynamics driven by uncertain input from the external, and as multi-scale dynamics which consists of slow memory dynamics and fast system dynamics. When the uncertain input sequences are modelled by stochastic processes, dynamics of learning is described by a random dynamical system. 
  In contrast to the traditional Fokker-Planck approaches \cite{chaudhari2017stochastic,mandt2017stochastic}, the random dynamical system approaches  enable the study not only of stationary distributions and global statistics, but also of the pathwise structure of stochastic dynamics. Based on non-autonomous and random dynamical system theory, it is possible to analyse stability and bifurcation in machine learning.

  We adopt a  {\it multi-layer perceptron}, a class of feed-forward neural networks, as the parametric model $f(x;\boldsymbol{\theta})$, where $\boldsymbol{\theta}$ is the parameter of neural networks and $x$ is the input data. 
  It is known that multi-layer perceptron  with a single hidden layer is a universal function approximator \cite{cybenko1989approximation}, and broadly used for solving generalisation problems.

  In recent years, behaviour of {\it stochastic gradient descent} \cite{robbins1951stochastic, amari1967theory} in finite settings, given by the following equation, has become more important: 
\begin{eqnarray}
	\boldsymbol{\theta}(t+1) &=& \boldsymbol{\theta}(t) - \frac{\eta}{S}\sum_{i=1}^S\nabla_{\boldsymbol{\theta}}l(x_i(t);\boldsymbol{\theta}(t)), 
	\label{eq:bgd}
\end{eqnarray}
where $\eta$ is the learning rate, $\{x_i(t)\}$ the input data, and $l(\cdot ;\boldsymbol{\theta})$ a loss function, typically given by the squared norm $||f(x;\boldsymbol{\theta})-T(x)||^2$, where $T(x)$ is the target function. 
For each time $t$, a finite set $\{x_i(t)\}_{i=1}^S$ of input data, called a {\it batch} of learning data, is randomly chosen, and Eq. \eqref{eq:bgd} is a stochastic dynamics with finite $S$.
  
  The {\it deterministic gradient descent} is given by the gradient dynamics of 
  the averaged potential  $E_x[l(x;\boldsymbol{\theta})]$; 
\begin{eqnarray}
  \boldsymbol{\theta}(t+1)&=&\boldsymbol{\theta}(t)-\eta \nabla_{\boldsymbol{\theta}}E_x[l(x;\boldsymbol{\theta})], 
    \label{eq:agd}
\end{eqnarray}
where $E_x[\cdot]$ denotes the expectation over $x$. 
The averaged potential $E_x[l(x;\boldsymbol{\theta})]$ is also known as the risk function. The approximation in Eq. \eqref{eq:agd} corresponds to the assumption that 
a large set of training data $x$ 
is given to the system at once, yielding the exact average potential $E_x[l(x;\boldsymbol{\theta})]$. Eq. \eqref{eq:agd} is a deterministic gradient dynamics.

\begin{figure}[htbp]
  \begin{center}
    \includegraphics[scale=0.7]{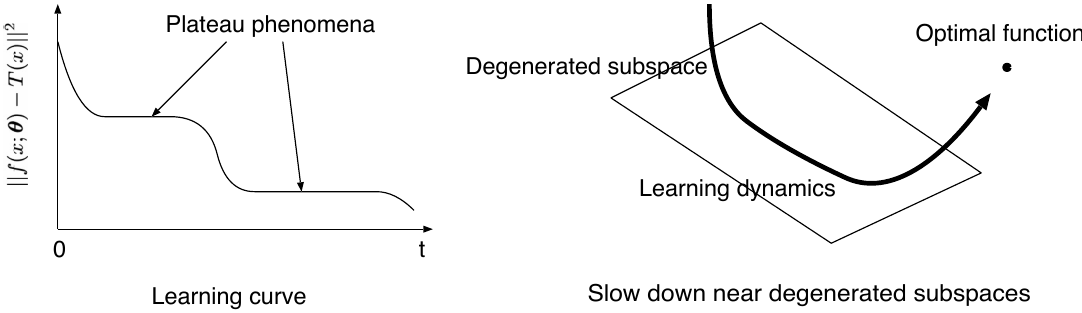}
  \end{center}
  \caption{A schematic view of the plateau phenomena (left) and a stagnant dynamics near the degenerated subspace (right). The dynamics of learning slows down near the attracting region in the degenerated subspace, but eventually escapes to the optimal.}
\label{fig:scheme}
\end{figure}

  In many cases, the gradient descent in multi-layer perceptron exhibits slow convergence to the optimal, known as a {\it plateau phenomenon} (see Fig. \ref{fig:scheme}), which makes the dynamics of learning slow down \cite{riegler1995online,huh2000online}. The plateau phenomenon is a chain of slow dynamics near attracting regions, during which the loss function reduces extremely small amount per unit time. Each plateau or trapping is caused by a saddle set or a Milnor attractor, which is an attractor with a zero-measure set of escape holes \cite{milnor1985concept}. In a degenerated subspace, in Eq. \eqref{eq:agd}, these attracting structures cause slow convergence to the optimal \cite{fukumizu2000local, cousseau2008dynamics, wei2008dynamics}. Fukumizu and Amari studied the deterministic gradient descent in the minimal three-layer perceptron with one input neuron, two neurons in the hidden layer, and one output neuron, which we call Fukumizu-Amari model, and found non-trivial plateau phenomena based on Milnor attractors \cite{fukumizu2000local}.

  Learning in multi-layer perceptrons with $S=1$, is called {\it online learning}. Online learning is modelled by a stochastic gradient descent dynamics with one-dimensional external force, which, however, is not well understood theoretically. Recently, plateau phenomena in online learning has been studied, based on the averaged dynamics $\bar{\theta}(t)$ of stochastic gradient descent with respect to  the deterministic potential $E_x[\nabla_{\boldsymbol{\bar{\theta}}}l(x;\boldsymbol{\bar{\theta}}(t))]$  by averaging the stochastic gradient over $x$ \cite{amari2018dynamics}. However, such an approximation is not fully valid in online learning, and recently a different approximation with stochastic dynamics near degenerated subspace is studied \cite{tsutsui2020center}.

  In order to elucidate the underlying mechanism of the plateau phenomena in multi-layer perceptrons, we here analyse stochastic gradient descent of Fukumizu-Amari model from a viewpoint of random dynamical systems without any averaging; 
\begin{eqnarray}
  \boldsymbol{\theta}(t+1)&=&\boldsymbol{\theta}(t)  -\eta \nabla_{\boldsymbol{\theta}}l(x(t);\boldsymbol{\theta}(t)), 
    \label{eq:sgd}
\end{eqnarray}
where $x(t)$ is a discrete time stochastic process. 
We show that, in a class of Fukumizu-Amari model, (1) attracting regions exist in multiply degenerated subspaces, (2) a strong plateau phenomenon emerges as a noise-induced synchronisation, which is not observed in deterministic gradient descent, (3) an optimal fluctuation size exists to minimise the escape time from the degenerated subspace.

    In the studies of stochastic gradient descent, the importance of the anisotropy of noise has been pointed out \cite{zhu2019anisotropic}.  A well-studied field is the Langevin gradient dynamics \cite{welling2011bayesian}, which is driven by isotropic noise, and is tractable by Fokker-Planck analysis; however, it is essentially different from the dynamics of stochastic gradient descent in the case of small batch size $S$. Keskar et al. found that stochastic gradient descent with small batch admits the convergence to flat minimisers \cite{hochreiter1997flat}, which results in better generalisation \cite{keskar2017large}. Some studies reported that small batch methods accelerate learning \cite{hoffer2017train, daneshmand2018escaping, xing2019walk}.
    
    In contrast, we find that there is an optimal fluctuation size which minimises the escape time. To the best of our knowledge, such a phenomenon has not been reported in studies on either deterministic or stochastic gradient descent.
    It is thus expected that our random dynamical system approach
    will shed new light on the analysis of online-learning.
    
The paper is organised as follows. In Section 2, we introduce the minimal three-layer perceptron, Fukimizu-Amari model, which shows plateau phenomena. In Section 3, we analyse stability of stochastic gradient descent in Fukumizu-Amari model; Global stability in the whole state space (Subsection 3.1) and local stability in the multiply degenerated subspace (Subsection 3.2) are discussed. In Section 4, we study phenomenology of ``noise-induced degeneration.'' In Section 5. we give a summary and an overview.

\newpage
\section{Stochastic gradient descent in multi-layer perceptrons}
The minimal model of multi-layer perceptrons which exhibits plateau phenomena is given by a three-layer perceptron with gradient descent \cite{fukumizu2000local}. 

\begin{figure}[htbp]
  \begin{center}
    \includegraphics[scale=0.6]{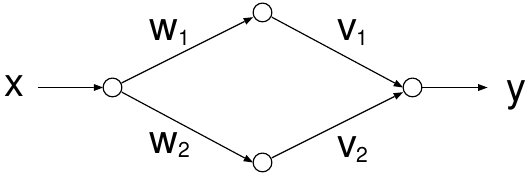}
    \end{center}
\caption{The three-layer perceptron: The nodes are  activation functions given by $\tanh(\cdot)$, 
and each edge indicates  a linear superposition with parameters $(w_1, w_2, v_1, v_2)$. The output $y$ is a function of the input $x$ and the parameters $(w_1, w_2, v_1, v_2)$.}
\label{fig:model}
\end{figure}

The equation of motion of the stochastic gradient descent \eqref{eq:sgd}, is explicitly written as follows;
\begin{eqnarray}
  \boldsymbol{\theta}(t+1)&=&\boldsymbol{\theta}(t)-\eta\nabla_{\boldsymbol{\theta}}l(x(t);\boldsymbol{\theta}(t)), ~(t=1,2,\ldots)
  \label{eq:dsgd}
\end{eqnarray}
where
\begin{eqnarray}
&&\boldsymbol{\theta}=(w_1,w_2,v_1,v_2) \in \boldsymbol{\Theta}\\
&&l(x;\boldsymbol{\theta})=\frac12(f(x;\boldsymbol{\theta})-T(x))^2\\
&&f(x;\boldsymbol{\theta})=v_1 \tanh(w_1 x)+v_2 \tanh(w_2 x). 
\end{eqnarray}
Here, $\boldsymbol{\Theta}$ is a domain of ${\bf R}^4$ called the parameter space, $x$ is an i.i.d. random variable subject to a probability distribution $\rho(x)$, $\eta\in [0,1]$ is the learning rate, and  $T(x)$ is the target function to be learnt. A Fukumizu-Amari model \cite{fukumizu2000local} is given by 
Eq. (\ref{eq:dsgd}) with a target function 
\begin{equation}
T(x)=2\tanh(x)-\tanh(4x), 
\label{eq:target}
\end{equation}
and $\rho(x)\sim N(0,\sigma^2)$. The optimal function $f(x;\boldsymbol{\theta}^*)$ with parameters $\boldsymbol{\theta}^*$ which minimise $l(x;\boldsymbol{\theta})$ is given by
\begin{eqnarray}
    \boldsymbol{\theta}^* &=&(1,4,2,-1), ~(-1,4,-2,-1), ~(1,-4,2,1), ~(-1,-4,-2,1),\nonumber\\ 
    & &
    (4,1,-1,2), ~(4,-1,-1,-2), ~(-4,1,1,2), ~(-4,-1,1,-2). 
\end{eqnarray}

Given Eqs. (4)-(8), the following subspaces  
\begin{equation}
  \boldsymbol{\theta}^{\dagger}=(w,w,v_1,2v-v_1), ~ (w,-w,v_1,v_1-2v), ~(w_1,w,0,2v), ~(w,w_2,2v,0)
  \label{eq:deg}
\end{equation}
define a class of degenerated functions
\begin{equation}
  f(x;\boldsymbol{\theta}^{\dagger})=2v\tanh(w x).
\end{equation}
Typically, the plateau phenomena emerge near the degenerated subspace \eqref{eq:deg} in the parameter space $\boldsymbol{\Theta}$ \cite{amari2018dynamics}.

    In this paper, we focus on the interplay between degenerated subspaces  $M_{w}=\{\boldsymbol{\theta}| w_1=w_2=w\}$ and  $M_{wv}=\{\boldsymbol{\theta}| w_1=w_2=w, v_1=v_2=v\}$. 
In the degenerated subspace $M_{w}$, the effective degrees of freedom of Eq. \eqref{eq:dsgd} decreases to $3$. Although we still have free variable $v_1$ in $M_w$, it does not contribute to a better function approximation. 
Furthermore, when a multiple degeneration to $M_{wv}$ occurs, the effective degrees of freedom decreases to $2$, and the dynamics cannot even return to the full parameter space $\boldsymbol{\Theta}$ by gradient descent.  

In many cases, the dynamics of learning stays near the degenerated subspace for a very long time. This trapping phenomenon is caused by neutral stability with vanishing gradients. 
In online learning, in addition to this neutrally stable trapping, ``stronger trapping'' based on multiple degeneration may occur as a noise-induced synchronisation \cite{pikovskii1984synchronization, teramae2004robustness,  sato2018analytical}. We call this type of degeneration as {\it noise-induced degeneration} in online learning.

\section{Stability analysis of stochastic gradient descent}
\subsection{Global attraction to the degenerated subspace}
We focus on the following random map derived from Fukumizu-Amari model (4)-(7);
\begin{eqnarray}
w_1(t+1)&=&w_1(t)- \eta x v_1(t)\frac{h(x;\boldsymbol{\theta},T)}{\cosh^2(w_1(t)x)},\\
w_2(t+1)&=&w_2(t)-\eta x v_2(t)\frac{h(x;\boldsymbol{\theta},T)}{\cosh^2(w_2(t)x)}, \\
v_1(t+1)&=&v_1(t)-\eta  \tanh(w_1(t)x)h(x;\boldsymbol{\theta},T),\\
v_2(t+1)&=&v_2(t)- \eta  \tanh(w_2(t)x)h(x;\boldsymbol{\theta},T),
\end{eqnarray}
where
\begin{equation}
  h(x;\boldsymbol{\theta},T)=v_1\tanh(w_1 x)+v_2\tanh(w_2 x)-T(x).
\end{equation}

The fluctuation $\sigma^2$ of the learning data is a control parameter. Our numerical experiments suggest that, for a broad region of large $\sigma^2$ and a positive measure set of initial conditions, and for a finite time, there exists attracting dynamics from full space $\boldsymbol{\Theta}$ to a degenerated subspace 
\begin{equation}
  M_{w}=\{\boldsymbol{\theta}~|~w_1=w_2=w\}. ~
\end{equation}
Furthermore, with large $\sigma^2$, we observe  attracting dynamics from the degenerated subspace $M_w$ to multiply degenerated subspace 
\begin{equation}
  M_{wv}=\{\boldsymbol{\theta}~|~w_1=w_2=w, ~v_1=v_2=v\}. 
\end{equation}

The attraction to $M_w$ is caused by a local minimum of the averaged potential $E_x[l(x;\boldsymbol{\theta})]$. 
Due to the valley formed  by steep gradients of the averaged potential, the gradient dynamics is dominant compared with the stochastic effects and approaches a neighbourhood of $M_w$ quickly (see Appendix A). 

To investigate local dynamics near the degenerated subspace $M_w$, one can introduce the following coordinate system 
\begin{equation}
	\left\{ \begin{array}{l}
		\displaystyle
		p = \frac{w_1+w_2}{2}, \quad
		q = \frac{v_1+v_2}{2}, \\[3mm]
		\displaystyle
		r = \frac{w_1-w_2}{2}, \quad
		s = \frac{v_1-v_2}{2},
	\end{array} \right.
	\label{eq:cdsys}
\end{equation}
and the corresponding transformed model 
\begin{eqnarray}
p(t+1)-p(t)&=&- \frac{\eta x h}{2} \left[\frac{q(t)+s(t)}{\cosh^2((p(t)+r(t))x)}+\frac{q(t)-s(t)}{\cosh^2((p(t)-r(t))x)}\right],\\
q(t+1)-q(t)&=&-\frac{\eta h}{2}\left[\tanh((p(t)+r(t))x)+\tanh((p(t)-r(t))x)\right],\\
r(t+1)-r(t)&=& -\frac{\eta x h}{2} \left[\frac{q(t)+s(t)}{\cosh^2((p(t)+r(t))x)}-\frac{(q(t)-s(t))}{\cosh^2((p(t)-r(t))x)}\right],\\
s(t+1)-s(t)&=&-\frac{\eta h}{2}\left[\tanh((p(t)+r(t))x)-\tanh((p(t)-r(t))x)\right],
\end{eqnarray}
where $h$ is given by 
\begin{equation}
  h=h(x;\boldsymbol{\theta}, T)=(q+s)\tanh((p+r)x)+(q-s)\tanh((p-r)x)-T(x).
\end{equation}

  In what follows, we confine ourselves to the situation where the parameters
  are in a bounded region and are stangnant away from the origin, which is for the synchronising dynamics typically observed in our numerical experiments.

  \begin{theorem}
    Assuming that
    \begin{enumerate}
    \item $\boldsymbol{\theta}$ is near $M_w$
    \item $\boldsymbol{\theta}$ is in a bounded region $D$, where 
      \[D:=\{\boldsymbol{\theta} | w_1,w_2\in [\kappa_1,\kappa_1+\delta], v_1,v_2\in [-\kappa_2, \kappa_2]\} ~~(0<\kappa_1, \kappa_2, \delta)\]
    \item The target function $T(x)$ is bounded as $|T(x)|\le L$,
    \end{enumerate}
 for a sufficiently small laerning rate $\eta>0$, the dynamics of $s(t)$ is contracting and approaching 0.
  \end{theorem}

  \begin{proof}
    A Taylor expansion of Eq. (23) around $r=0$ yields
\begin{equation}
  s(t+1)
=\eta C_1+\left[1-\eta C_2 \right]s(t)+O(r^3)
        \label{eq:ss0}
\end{equation}
where
\begin{eqnarray}
C_1&=& -\frac{rx (2q\tanh(px)-T(x))}{\cosh^2(px)}\\
C_2 &=& \frac{ 2r^2 x^2}{\cosh^4(px)}. 
\end{eqnarray}

Since $|\frac{x}{\cosh^2 x}|<1$ and $p\simeq w_1\simeq w_2$ near $M_w$, we have 
\begin{equation}
  0\le |C_1|\le \frac{|r x|(2|q|+L)}{\cosh^2(px)}\le \frac{\kappa_2(2\kappa_2+L) |x |}{\cosh^2(\kappa_1 x)} < \frac{ \kappa_2(2\kappa_2+L)}{\kappa_1}
\end{equation}
and
\begin{equation}
   0\le C_2\le \frac{ 2\kappa_2^2 x^2}{\cosh^4(\kappa_1 x)} < \frac{ 2\kappa_2^2}{\kappa_1^2}.
\end{equation}
Consequently, for sufficiently small $r$ and $\eta$, we have 
\begin{equation}
  s(t+1)
\simeq \epsilon+ c s(t), 
        \label{eq:ss}
\end{equation}
where $\epsilon=\eta C_1\simeq 0$, $0<c=1-\eta C_2<1$. 
  \end{proof}

  The theorem 1 implies existence of the second global attraction from the neighbourhood $M_w$ to a neighbourhood of $M_{wv}$.  If $s(t)$ is exactly $0$, we observe a total synchronisation with $w_1=w_2$ and $v_1=v_2$ on the degenerated subspace $M_{wv}$. In practice, due to the external inputs $x$, $s(t)$ approaches to $0$ and fluctuates near $0$.


  The multiple degeneration from $M_w$ to $M_{wv}$ is a characteristic behaviour of the stochastic gradient descent dynamics. In the case of the deterministic gradient descent dynamics, the dynamics near $M_{w}$ in the direction of $s$ is neutral because $r$ converges to $0$ monotonically \cite{tsutsui2020center}. 
By the contrary, in the stochastic gradient descent dynamics, the dynamics of 
$s$ can be contracting because $r$ fluctuates around $0$ \cite{tsutsui2020center}. This phenomenon is comparable with noise-induced synchronisation in random dynamical systems. A typical example of synchronisation is given by uncoupled phase oscillators driven by common noise (see Appendix D). It is known that the Lyapunov exponent of the stochastic phase oscillator is negative while those of the deterministic dynamics is zero \cite{teramae2004robustness, sato2018dynamical}.

\vspace{2mm}
In the next section, we show that there exists yet another noise-induced trapping mechanism in the multiply degenerated subspace $M_{wv}$.

\subsection{Local attraction in the degenerated subspace}

The equation of motion in $M_{wv}$ is given by the following two-dimensional random map of $w(=w_1=w_2)$ and  $v(=v_1=v_2)$;

\begin{eqnarray}
   w(t+1)&=&w(t)-\eta x v(t)\frac{2v(t)\tanh(w(t)x)-T(x)}{\cosh^2(w(t)x)},\\
   v(t+1)&=&v(t)-\eta\tanh(w(t)x)[2v(t)\tanh(w(t)x)-T(x)].
  \label{eq:2ddsdg}
\end{eqnarray}
or equivalently
\begin{eqnarray}
\begin{pmatrix} w(t+1)-w(t) \\ v(t+1)-v(t) \end{pmatrix}
=\eta \cdot g(x; w, v, T(x)),
\end{eqnarray}
where
\begin{equation}
 g(x; w, v, T) =- [2v\tanh(wx)-T(x)]
 \begin{pmatrix} \displaystyle\frac{vx}{\cosh^2(wx)} \\ \\ \tanh(wx)\end{pmatrix}.
\end{equation}

 \begin{theorem}
   If $T(x)\neq 2v\tanh(wx)$, for a sufficiently small $\eta>0$, a point $(w,v)$ in the dynamics (31), (32) is either an attracting point or a saddle point. 
 \end{theorem}
 
\begin{proof}
The Jacobian matrix of $g$ is given by 
\begin{equation}
  J(x;w,v)=
    \left[
    \begin{array}{cc}
      -\frac{2vx^2\left[T(x)\tanh(wx)-3v \tanh^2(wx)+v\right]}{\cosh^2(wx)}
      &\frac{x \left[T(x)-vq\tanh(wx)\right]}{\cosh^2(wx)}\\
      \frac{x \left[T(x)-4v\tanh(wx)\right]}{\cosh^2(wx)}
        &-2\tanh^2(wx)
    \end{array}
    \right],   
\end{equation}
or equivalently 
\begin{eqnarray}
  J(x;w,v)&=&-2 \left(\begin{array}{c} \frac{vx}{\cosh^2(wx)}\\ \tanh(wx) \end{array}\right)
    \left(\begin{array}{cc} \frac{vx}{\cosh^2(wx)} &\tanh(wx) \end{array}\right) \nonumber\\
    &&-\left(2v\tanh(wx)-T(x)\right)
    \left(\begin{array}{cc}
      -\frac{2vx^2\tanh(wx)}{\cosh^2(wx)} &\frac{x}{\cosh^2(wx)}\\
      \frac{x}{\cosh^2(wx)} &0
    \end{array}\right).
    \label{eq:j2}
\end{eqnarray}
Let the eigenvalues of $J$ at a point $(w,v)$ be $\mu_-(x;w,v)$ and $\mu_+(x;w,v)$ ($\mu_-\leq\mu_+$) (See Appendix B).
Since 
\begin{equation}
 \det \left(\begin{array}{cc}
      -\frac{2vx^2\tanh(wx)}{\cosh^2(wx)} &\frac{x}{\cosh^2(wx)}\\
      \frac{x}{\cosh^2(wx)} &0
    \end{array}\right)
    =- \left(\frac{x}{\cosh^2(px)}\right)^2 < 0
\end{equation}
for $x\neq 0$, the second term of Eq. \eqref{eq:j2} has both positive and negative eigenvalues if  $2v\tanh(wx)-T(x)\neq 0$. The first term of Eq. \eqref{eq:j2} is negative semidefinite. 
Thus, $\mu_-(x;w,v)$ is negative whenever $2v\tanh(wx)-T(x)\neq 0$. Therefore, for  sufficiently small $\eta$, the point $(w,v)$ is attracting when $\mu_+(x;w,v)$ is non-positive;   otherwise, it is a saddle.
\end{proof}

The theorem 2 implies that depending on $x$, each point in $M_{wv}$ may be weakly attracting. The dynamics near $(w,v)$ on $M_{wv}$ is characterised by the sign of $\mu_+(x;w,v)$ as long as $\eta$ is small.

\begin{figure}[htbp]
  \begin{center}
    \includegraphics[scale=0.53]{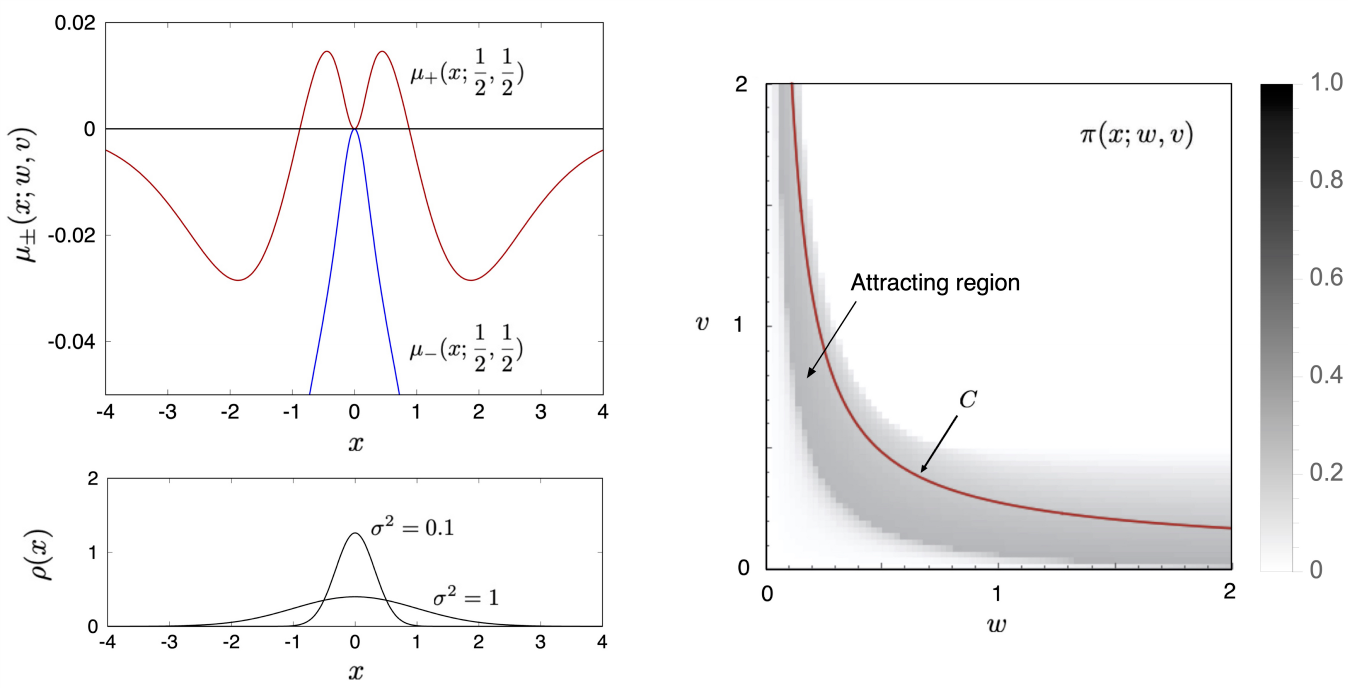}
  \end{center}
\caption{(Left) The eigenvalues $\mu_+(x;1/2,1/2)$ (red), and $\mu_-(x;1/2,1/2)$ (blue) of $J(x;1/2,1/2)$ as well as the distribution $\rho(x)$ are depicted as functions of $x$. The parameters are set as $T(x)=2\tanh(x)-\tanh(4x)$, $\eta=0.1$, and $\sigma^2$=0.1,1. When the fluctuation $\sigma^2$ is small, $x$ is frequently sampled near zero, and the point $(w,v)=(1/2,1/2)$ is a saddle point; otherwise, it is an attracting point. (right) The probability $\pi(x;w,v)={\rm Prob}[\mu_+(x;w,v)\leq 0]$ for $\sigma^2=1$ plotted in $[0,2]^2$ on $M_{wv}$. The red curve $C$ indicates the valley formed by steep gradients of the averaged potential (see Appendix A). }
\label{fig:stability}
\end{figure}

As an example, we investigate the dynamics near $(w,v)=(1/2,1/2)$ by computing the eigenvalues $\mu_{\pm}(x;1/2,1/2)$  of $J(x;w,v)$ explicitly.  We see from (Fig.\ref{fig:stability} (left)) that 
the dynamics is attracted to the point $(w,v)=(1/2,1/2)$  when the fluctuation $\sigma^2$ is sufficiently large. Put differently, the strong noise may ``stabilise'' the dynamics near $(w,v)$ in $M_{wv}$. As a result, the residual time near $M_{wv}$ is extended and a stronger plateau phenomenon is observed.

Fig.\ref{fig:stability} (right) shows  
the numerically computed probability distribution  
\begin{equation}
\pi(x;w,v)={\rm Prob}[ \mu_+(x;w,v)\leq 0], 
\end{equation}
with $\sigma^2=1$, where higher values of $\pi(x;w,v)$ corresponds to the darker tones. The red curve $C$ indicates the approximated one-dimensional valley formed by steep gradients of the averaged potential. In this case, $C$ includes a local minimum of the averaged potential. The dark grey region, where $\pi(x;w,v)$ is large, may cause stronger trapping dynamics on $M_{wv}$. 
To the contrary, with smaller fluctuation $\sigma^2=0.1$, the grey attracting region in Fig. \ref{fig:stability} (right) disappears because most points in $M_{wv}$ are saddle points. Thus, there exists another type of noise-induced phenomenon, i.e., the emergence of an attracting region near $M_{wv}$ due to the large fluctuation $\sigma^2$, resulting in a substantial extension of the escape time from $M_{wv}$. In the next section, we discuss the escape time from $M_{w,v}$ phenomenologically.

\section{Noise-induced degeneration}

Summarising the main results, noise-induced degeneration and plateau phenomena emerge through the following three processes;
\begin{enumerate}
    \item {The first global attraction to a neighbourhood of the degenerated subspace $M_w$. ~(Appendix A)}
    \item{The second global attraction from the neighbourood $M_w$ to a neighbourhood of the multiply degenerated subspace $M_{wv}$. ~(Theorem 1)}
    \item{The third weak attraction, by the attracting regions in the neighbourhood of $M_{wv}$. ~(Theorem 2)} 
\end{enumerate}

To examine the above three-step scenario, we made numerical experiments to compute the global dynamics and the escape time from the multiply degenerated subspace.

\subsection{Global dynamical phenomena}
The global dynamics of stochastic gradient descent in $\boldsymbol{\Theta}$ is shown in Fig.\ref{fig:sim}. It depicts converging dynamics to pullback attractors \cite{sato2020} (see Appendix C) in Fukumizu-Amari model with $T(x)=2\tanh(x)-\tanh(4x)$, $\eta=0.1$, and $\sigma^2=0.1, 1.0$. The red and blue dots represent paths of $(w_1,w_2)$ and $(v_1,v_2)$,  respectively, starting from different initial conditions. Both dynamics are plotted together in each panel. The grey points correspond to the optimal attractors $\boldsymbol{\theta}^*$.  The degenerated subspaces $w_1=w_2$ and $v_1=v_2$ are depicted as a single line. In the case of $\sigma^2=1.0$, clear noise-induced degeneration, i.e. $|w_1-w_2|\rightarrow 0$ near $M_w$ at $\tau=10000$, followed by $|v_1-v_2|\rightarrow 0$ near $M_{wv}$ at $\tau=30000$, are observed. Due to this type of noise-induced degeneration, about $20$ percentage of the sample paths stay near the attracting region in $M_{wv}$ for an extremely long time, which is shown in a dashed circle at $\tau=100000$. 

\begin{figure}[htbp]
  \begin{center}
    \includegraphics[scale=0.9]{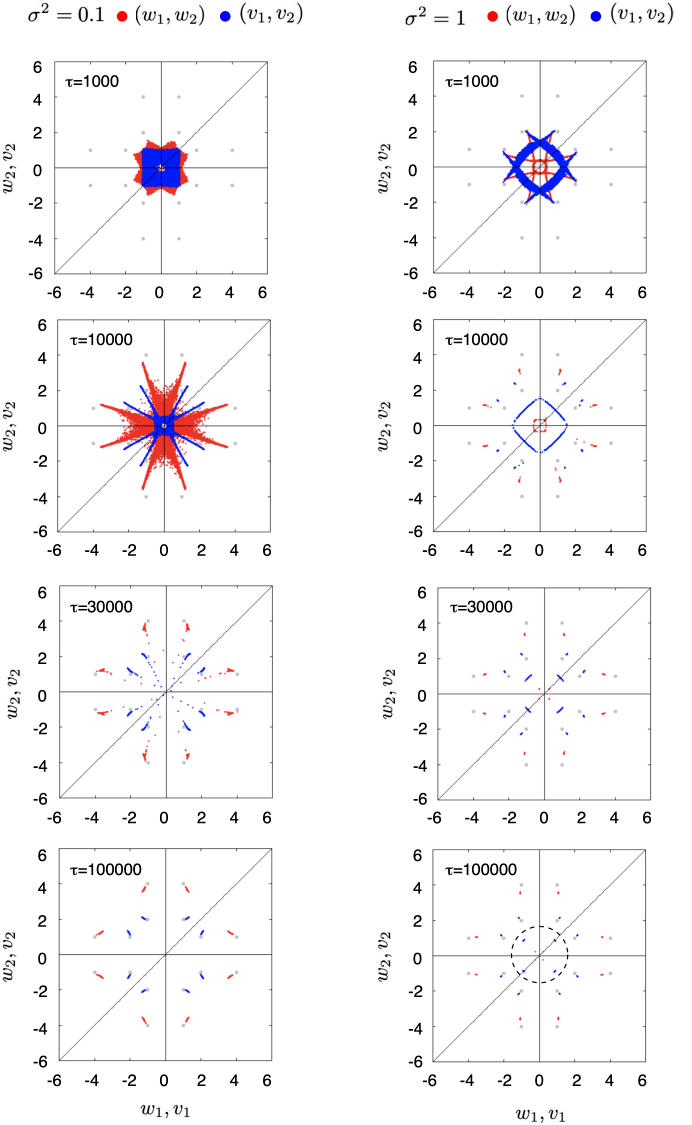}
  \end{center}
  \caption{Finite time pullback attractors (see appendix C) with the pullback time $\tau=1 000$, $\tau=10 000$, $\tau=30 000$, and $\tau=100 000$ in the full space $\boldsymbol{\Theta}$. 
  Parameters are set as $T(x)=2\tanh(x)-\tanh(4x)$, $\eta=0.1$, and $\sigma^2=0.1$ (left) and $\sigma^2=1.0$ (right). The red and blue dots represent paths of $(w_1,w_2)$ and $(v_1,v_2)$,  respectively, starting from different initial conditions. Both dynamics are plotted together in each panel.  
  The grey points correspond to the optimal attractors $\boldsymbol{\theta}^*$.  The degenerated subspace $w_1=w_2$ and $v_1=v_2$ are depicted as a single line. 
  A typical noise realisation $\{x\}$ is fixed and dynamics is developed with $10^5$ different initial conditions $\boldsymbol{\theta}(0)\in[-1,1]^4$. When $\sigma^2=1.0$, the trapping dynamics near $M_{wv}$ is observed in the attracting region indicated by a dashed circle. }
\label{fig:sim}
\end{figure}

\subsection{Escape time from the degenerated subspace}
We numerically compute the average escape time from the attracting region on $M_{wv} ~(w,v>0)$ for several values of $\sigma^2$. The escape time from the region $[-2,2]\times[-2,2]$, which includes the multiply degenerated subspace $M_{wv} ~(w,v>0)$, is averaged over $10^5$ initial conditions and $10^3$ noise realisations.
In Fig. 5, the average escape time $\tau^*$ as a function of the fluctuation size $\sigma^2$ is depicted in log-log plot. 
One can see that larger fluctuation size induces longer escape time because of noise-induced degeneration becoming effective around $\sigma^2=0.15$.
When noise is weak, around $\sigma^2=0.004$, $\tau^*$ is large, and when noise is strong, around $\sigma^2=0.15$, $\tau^*$ is large again. 
In the intermediate scale, an optimal fluctuation size around $\sigma^2\simeq 0.07$ exists to minimise the escape time from the degenerated subspace. 
Thus, the classical Kramer's escape view based on Brownian particle in a potential is insufficient, and the pathwise analysis in terms of random dynamical systems theory is needed.

\begin{figure}[htbp]
  \begin{center}
    \includegraphics[scale=0.8]{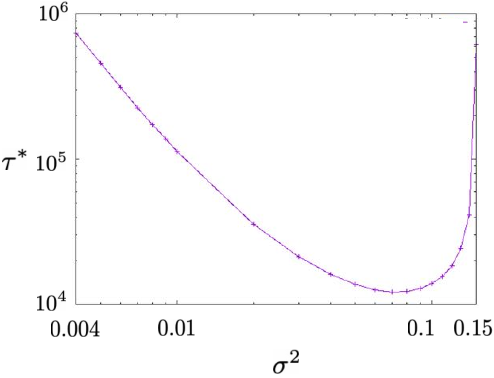}
  \end{center}
  \caption{The averaged escape time from $[-2,2]\times[-2,2]$, $\tau^*$ as a function of $\sigma^2$ in log-log plot. The other parameters are same as in the numerical computation in Fig. 4. Stronger noise induces longer escape time because of noise-induced degeneration. An optimal fluctuation size is $\sigma^2\simeq 0.07$.}
  \label{fig:esc}
\end{figure}

\subsection{Random attractors in stochastic gradient descent}
If there exists an attractor $A$ on a subspace $M$, and $A$ is not an attractor in the full space $\boldsymbol{\Theta}$, it is called a relative attractor in $\boldsymbol{\Theta}$ \cite{skufca2006edge}. Thus, if there exists an attractor on $M_{wv}$, it is a relative attractor. If there is no random compact invariant set in a random dynamical system, all random attractors are random Milnor attractors \cite{ashwin1999minimal}. 
Our conjecture is that both relative attractors (or attracting region) in the multiply degenerated subspaces $M_{wv}$ and the optimal attractors $\boldsymbol{\Theta}$ are random Milnor attractors. The dynamics of learning can come and go with each attractor in a long time scale.

In general, non-optimal stable random attractors may exist in stochastic gradient descent learning with small $\eta$, because $T(x)$, $f(x;\boldsymbol{\theta})$, and $\nabla_{\boldsymbol{\theta}}l(x;\boldsymbol{\theta})$ are bounded for any $\rho(x)$. In these cases, if an initial point $\boldsymbol{\theta}(0)$ does not belong to the effective basins of the optimal attractors $\boldsymbol{\theta}^*$, the attracting region near $\boldsymbol{\theta}^{\dagger}$ may become a stable random attractor, which is completely separated from the optimal attractor $\boldsymbol{\theta}^*$, and the orbits can stay there for an arbitrarily long time.

\subsection{An extension to larger networks and applications to practical problems}
Our results can be extended to a larger three-layer perceptron with $N$ hidden-layer neurons and $L$ dimensional output neurons. We consider a parameterized function
\[ f_i(x;\boldsymbol{\theta})=\sum_{k=1}^N v_k\tanh(w_k x) ~~(i=1,\ldots,L)\]
where 
\[\boldsymbol{\theta}=(w_1,\ldots,w_N, v_1\ldots,v_N)\].
Near the subspaces defined by 
\[w_1=\ldots=w_j=w_a, ~~w_{j+1}=\ldots=w_M=w_b ~~(j=1,\ldots,M, ~1<M<N),\]
\[v_1=\ldots=v_j=v_a, ~~v_{j+1}=\ldots=v_M=v_b ~~(j=1,\ldots,M, ~1<M<N),\]
we have a degenerated funciton
\begin{eqnarray}
f_i(x;\boldsymbol{\theta})&=&jv_a\tanh(w_a x)+(M-j)v_b\tanh(w_b x)
+\sum_{k=M+1}^N v_k\tanh(w_k x) \nonumber\\
&=& Mg(x;w_a,w_b,v_a,v_b)+\sum_{k=M+1}^N v_k\tanh(w_k x),
\end{eqnarray}
where $i=1,\ldots,L, ~j=1,\ldots M$, and
\[g(x;w_a,w,_b,v_a,v_b)=\frac{j}{M}v_a\tanh(w_a x)+\frac{M-j}{M}v_b\tanh(w_b x).\]
Thus, noise-induced degeneration in learning the function $g(x;w_a,w_b,v_a,v_b)$ occurs again in the same way in the subspace $(w_a,w_b,v_a,v_b)$. We have many of such degenerated subspaces hierarchically embedded in the entire parameter space $\boldsymbol{\Theta}$. Indeed, it is known that these inevitable singular structure is a generic property of all types of neural networks \cite{amari2018dynamics}.

From the viewpoint of nonlinear physics, the plateau phenomena, noise-induced degeneration, and noise-induced synchronization between two variables $(w_1,v_1)$ and $(w_2,v_2)$ are all understood as same dynamical structure. Since synchronisation is an universal phenomenon, they are expected to be observed in a broad class of dynamical systems, and so as in a broad class of online learning. 
If the system has a certain local symmetry, synchronisation almost always occurs in that locally symmetric space, 
and the presented trapping phenomena near the multiply degenerated subspace may occur and possibly affect global dynamics. In some cases, the state space degeneration is rather useful, because, if there is no degeneration, we faces to problems of over-fitting, which is caused by explosion of the effective degrees of freedom. Thus, it is important to study the trade-off between degeneration and explosion of the effective degree of freedom in terms of nonlinear physics.

In practice, unless the learning requires the full dimension of the parameters, we need degeneration to some extent to avoid over-fitting, however that causes the possible strong plateau phenomena in the learnign dynamics. We may choose an "optimal fluctuation size'' to sample the input data $x$, which is examplified in Fig. 5, and which, however in general, depends on the complexity of problems and models.

\section{Conclusion}
In Fukumizu-Amari model, there exist characteristic fluctuation sizes of the training data, with which the dynamics shows strong plateau phenomena.  Starting from an initial point in the full space, the dynamics of learning is attracted by a degenerated subspace by the gradient, and then, is attracted by a multiply degenerated subspace by noise-induced synchronisation.
We called the second attraction noise-induced degeneration, which is not observed in deterministic gradient descent. Furthermore, when noise is strong, near the multiply degenerated subspace, attracting regions emerge and the residual time near there can be extremely long. 

In the attracting region in the multiply degenerated subspace, an optimal fluctuation size exists to minimises the escape time, and stronger noise induces longer escape time. This implies that the classical Kramer's escape view is insufficient, and the pathwise analysis in terms of random dynamical systems theory is needed.

Noise-induced degeneration presented in this paper is expected to be one of the key elements to understand advantages of variety of machine learning to control the appropriate size of the degrees of freedom. As pointed out in \cite{amari2018dynamics}, the degeneration may be weaken in perceptrons with many hidden layers, which could explain in part the advantage of deep learning. We here add an opposite possibility: the existence of many hidden layer may result in stronger degeneration, which could explain the reason why the over-parameterised neural networks work well. 

Further studies on stability of random attractors, geometry of random basins, and bifurcations of stochastic gradient descent of larger neural networks will be reported elsewhere. Our approach would shed new light on various problems in machine learning from a viewpoint of random dynamical systems theory. 

\section{Acknowledgements}
          The authors thank the anonymous referees for meaningful comments. YS is supported by the external fellowship of London Mathematical Laboratory and the Grant in Aid for Scientific Research (C) No. 18K03441, JSPS. 
        Authors are supported by the Grant in Aid for Scientific Research (B) No. 17H02861, JSPS. 

\begin{appendix}
\section{Local minima in the averaged dynamics}
In this paper, we have treated $\tanh(\cdot)$ as an activation function. However, in order to analyse the averaged potential, we herein use the function $h(u):=\text{erf}(x/\sqrt{2}):=\sqrt{2/\pi}\int_0^x e^{-t^2/2} dt$ as an approximation of $\tanh(\cdot)$ to understand the qualitative behaviour of the dynamics. 
According to \cite{biehl1995learning}, for a network 
\begin{equation}
    f(x;\boldsymbol{\theta}) = v_1 h(w_1 x) + v_2 h(w_2 x), 
\end{equation}
and a target function which is denoted as
\begin{equation}
    T(x) = \nu_1 h(\omega_1 x)+\nu_2 h(\omega_2 x), 
\end{equation}
for some values $\nu_1,\nu_2,\omega_1,\omega_2\in\boldmath{R}$, 
the averaged potential is given by
\begin{eqnarray}
    L(\boldsymbol{\theta})&=&E_x\left[ \frac{1}{2}\left(f(x;\boldsymbol{\theta})-T(x)\right)^2 \right] \notag\\
	&=& \frac{1}{\pi} 
	    \sum_{i,j=1}^2 v_i v_j \Phi(w_i,w_j) 
		- \frac{2}{\pi}\sum_{i,a=1}^2 v_i\nu_a \Phi(w_i,\omega_a) 
		+ \text{const}, \\
	\Phi(\zeta_1,\zeta_2)
	&:=& \arcsin\left(\frac{\sigma^2\zeta_1\zeta_2}{\sqrt{1+\sigma^2\zeta_1^2}\sqrt{1+\sigma^2\zeta_2^2}}\right). 
\end{eqnarray}

When $w_1\neq 0$ and $w_2\neq 0$, the function $L(\boldsymbol{\theta})$ is a quadratic in  $(v_1,v_2)$, and thus has a minimiser. 
In particular, when $w_1=w_2=w$ is fixed and $v_1=v_2=v$, 
\begin{equation}
    L^*(w,v):= L(\boldsymbol{\theta}) = \frac{4}{\pi} \Biggl( 
	    v^2 \Phi(w,w) 
		- v \sum_{a=1}^2 \nu_a \Phi(w,\omega_a) 
		\Biggr)
		+ \text{const.}
\end{equation}
takes minimum at
\begin{equation}
	v^*(w;\sigma^2) = \frac{\sum_a \nu_a \Phi(w,\omega_a)}{2\Phi(w,w)}.  
\end{equation}
Hence, a minimiser of $L^*$ lies in the one-dimensional valley $\{ (w, v^*(w;\sigma^2)) | w\in\boldmath{R} \}$.

For $(\omega_1,\omega_2,\nu_1,\nu_2)=(1,4,2,-1)$ and $\sigma^2=1$, in particular, we have the red curve $C$ in Fig. \ref{fig:stability} (right) as 
\begin{equation}
	v^*(w;1) = \frac{
	\left(
	    2\arcsin\left(\frac{w}{\sqrt{2}\sqrt{1+w^2}}\right)
	    -\arcsin\left(\frac{4w}{\sqrt{17}\sqrt{1+w^2}}\right)
	\right)}{\left(2\arcsin\left(\frac{w^2}{1+w^2}\right)\right)}.   
	\label{eq:min}
\end{equation}

\section{Eigenvalues of Jacobian of the dynamics in  the multiply degenerated subspace}

The Jacobian of $g$ on the multiply degenerated subspace $M_{wv}$ is given by 
\begin{equation}
    J=
    \left[
    \begin{array}{cc}
      -\frac{2 vx^2\left[T(x)\tanh(wx)-3v \tanh^2(wx)+v\right]}{\cosh^2(wx)}
      &\frac{ x \left[T(x)-vq\tanh(wx)\right]}{\cosh^2(wx)}\\
      \frac{ x \left[T(x)-4v\tanh(wx)\right]}{\cosh^2(wx)}
        &-2\tanh^2(wx)
    \end{array}
    \right]. 
\end{equation}
The eigenvalues of $J$ are given by
\begin{equation}
 \mu_{\pm}(x;w,v)=A\pm \frac{1}{8\cosh^4(wx)}\sqrt{B},
\end{equation}
where
\begin{eqnarray}
 A&=&-1+\frac{1+2v^2x^2-vx^2\tanh(wx) T(x)}{\cosh^2(wx)}-\frac{3v^2x^2}{\cosh^4(wx)},\\
 B&=&32 x^2 \cosh^2(w x)C+ \nonumber\\
 &&\hspace{-20mm}\left(-1+16 v^2 x^2-8 v^2 x^2 \cosh(2 w x)+\cosh(4 w x) +4 v x^2 \sinh(2 w x) T(x)\right)^2\\
 C&=&8 v^2 (2+\cosh(2 w x)) \sinh^2(w x) -v (6 \sinh(2 w x)\nonumber\\
 &&+\sinh(4 w x)) T(x)+2 \cosh^2(w x) T(x)^2.
\end{eqnarray}

\section{Pullback attractors in random dynamical systems}

Let $\theta$ acts on the probabilistic space of noise realisations $\Omega$, and $\theta_t \omega$ is the path taken at time $t$ by the noise realisation $\omega\in\Omega$. The random dynamical system is represented by the pair $(\theta, \phi)$, where $\phi$ denotes the dynamics in the state space $X$, driven by a noise realisation $\theta_t \omega$. The {\it pullback attractor} $A(t,\omega)$ of a random dynamical system is defined as a random invariant set of $X$ that satisfies
\begin{equation}
\lim_{\tau\rightarrow\infty} \mbox{dist}(\phi(\tau, \theta_{t-\tau}\omega) B, A(t,\omega))=0,
\label{eq:PBA}
\end{equation}
for any bounded set $B \subset X$, where ${\rm {dist}} (C, D)$ denotes the Hausdorff distance between two subsets $C$ and $D$ of $X$ \cite{chekroun2011stochastic}. 
 
We call the following $\tau$-pullback image of $B$ as {\it finite time pullback attractor} or {\it $\tau$-pullback attractor}, which is given by 
 \begin{equation}\label{eq:snapshot}
\widetilde{A}_{\tau}^B(t,\omega) = \phi(\tau, \theta_{t-\tau}\omega) B, 
\end{equation}
where $\tau$ is called pullback time. For a given $\tau$, the set $\widetilde{A}_{\tau}^B(t,\omega)$ represents a finite space-time structure, which may include transient orbits and densities. Each invariant sets in Fig. \ref{fig:sim} are finite time pullback attractors with pullback time $\tau$.

\section{Noise-induced synchronisation}
A stochastic phase oscillator is given by 
\begin{equation}
    d\phi=\omega dt+ \sin\phi \circ dW_t, 
    \label{eq:spo}
\end{equation}
in Stratonovich form, where $\omega$ is a constant, $\phi\in(0,2\pi]$ is phase on circle, and $W_t$ is  the Wiener process with $dW_t \sim N(0,\sigma^2)$. 
The linearisation along a fixed solution $\phi$ is given by 
\begin{equation}
    d\psi=\cos\phi\cdot\psi\circ dW_t, 
    \label{eq:spod}
\end{equation}
Let $r=\log |\psi|$, then we have 
\begin{equation}
    dr=\cos\phi\circ dW_t, 
    \label{eq:spodr}
\end{equation}
or, equivalently in Ito form, 
\begin{equation}
dr
=-\frac{\sigma^2}{2}\sin^2\phi dt+\cos\phi dW_t. 
\end{equation}
Thus, the Lyapunov exponent $\lambda$ of \eqref{eq:spo} is given by
\begin{equation}
     \lambda=\lim_{T\rightarrow\infty}\frac{r(T)}{T}
     =-\lim_{T\rightarrow\infty}\frac{1}{T}\int_0^T 
     \frac{\sigma^2}{2}\sin^2\phi   dt.
\end{equation}
Assuming that the fluctuation $\sigma^2$ is small, the dynamics is ergodic, and the invariant density is approximately uniform on circle, we have 
\begin{equation}
     \lambda
     \simeq -\frac{1}{2\pi}\int_0^{2\pi} \frac{\sigma^2}{2} \sin^2\phi  d\phi 
     =-\frac{\sigma^2}{4}<0.
\end{equation}

\end{appendix}

\bibliographystyle{plain}
\bibliography{rds}

\begin{thebibliography}{10}

\bibitem{amari1967theory}
Shun-ichi Amari.
\newblock A theory of adaptive pattern classifiers.
\newblock {\em IEEE Transactions on Electronic Computers}, 3:299--307, 1967.

\bibitem{amari2018dynamics}
Shun-ichi Amari, Tomoko Ozeki, Ryo Karakida, Yuki Yoshida, and Masato Okada.
\newblock Dynamics of learning in mlp: Natural gradient and singularity
  revisited.
\newblock {\em Neural computation}, 30(1):1--33, 2018.

\bibitem{ashwin1999minimal}
Peter Ashwin.
\newblock Minimal attractors and bifurcations of random dynamical systems.
\newblock {\em Proceedings of the Royal Society of London. Series A:
  Mathematical, Physical and Engineering Sciences}, 455(1987):2615--2634, 1999.

\bibitem{biehl1995learning}
Michael Biehl and Holm Schwarze.
\newblock Learning by online gradient descent.
\newblock {\em Journal of Physics A}, 28:643--656, 1995.

\bibitem{chaudhari2017stochastic}
Pratic Chaudhari and Stefano Soatto.
\newblock Stochastic gradient descent performs variational inference, converges
  to limit cycles for deep networks.
\newblock {\em arXiv preprint arXiv:1710.11029}, 2017.

\bibitem{chekroun2011stochastic}
Micka{\"e}l~D Chekroun, Eric Simonnet, and Michael Ghil.
\newblock Stochastic climate dynamics: Random attractors and time-dependent
  invariant measures.
\newblock {\em Physica D: Nonlinear Phenomena}, 240(21):1685--1700, 2011.

\bibitem{cousseau2008dynamics}
Florent Cousseau, Tomoko Ozeki, and Shun-ichi Amari.
\newblock Dynamics of learning in multilayer perceptrons near singularities.
\newblock {\em IEEE Transactions on Neural Networks}, 19:1313--1328, 2008.

\bibitem{cybenko1989approximation}
George Cybenko.
\newblock Approximation by superpositions of a sigmoidal function.
\newblock {\em Mathematics of control, signals and systems}, 2(4):303--314,
  1989.

\bibitem{daneshmand2018escaping}
Hadi Daneshmand, Jonas Kohler, Aurelien Lucchi, and Thomas Hofmann.
\newblock Escaping saddles with stochastic gradients.
\newblock {\em arXiv preprint arXiv:1803.05999}, 2018.

\bibitem{fukumizu2000local}
Kenji Fukumizu and Shun-ichi Amari.
\newblock Local minima and plateaus in hierarchical structures of multilayer
  perceptrons.
\newblock {\em Neural networks}, 13(3):317--327, 2000.

\bibitem{hochreiter1997flat}
Sepp Hochreiter and J\u{u}rgen Schmidhuber.
\newblock Flat minima.
\newblock {\em Neural Computation}, 9:1--42, 1997.

\bibitem{hoffer2017train}
Elad Hoffer, Itay Hubara, and Daniel Soudry.
\newblock Train longer, generalize better: Closing the generalization gap in
  large batch training of neural networks.
\newblock {\em In NIPS}, 30:1731--1741, 2017.

\bibitem{huh2000online}
N-J Huh, J-H Oh, and K~Kang.
\newblock On-line learning of a mixture-of-experts neural network.
\newblock {\em Journal of Physics A, Mathematical and General}, 33:8663--8672,
  2000.

\bibitem{keskar2017large}
Nitish~Shirish Keskar, Dheevatsa Mudigere, Jorge Nocedal, Mikhail Smelyanskiy,
  and Ping Tak~Peter Tang.
\newblock On large-batch training for deep learning: Generalization gap and
  sharp minima.
\newblock {\em In ICLR}, 2017.

\bibitem{mandt2017stochastic}
Stephan Mandt, Matthew~D Hoffman, and David~M Blei.
\newblock Stochastic gradient descent as approximate bayesian inference.
\newblock {\em Journal of Machine Learning Research}, 18:4873--4907, 2017.

\bibitem{milnor1985concept}
John Milnor.
\newblock On the concept of attractor.
\newblock In {\em The theory of chaotic attractors}, pages 243--264. Springer,
  1985.

\bibitem{pikovskii1984synchronization}
A~S Pikovskii.
\newblock Synchronization and stochastization of array of self-excited
  oscillators by external noise.
\newblock {\em Radiophysics and Quantum Electronics}, 27(5):390--395, 1984.

\bibitem{riegler1995online}
Peter Riegler and Biehl Michael.
\newblock On-line backpropagation in two-layered neural networks.
\newblock {\em Journal of Physics A, Mathematical and General}, 28:L507--L513,
  1995.

\bibitem{robbins1951stochastic}
Herbert Robbins and Sutton Monro.
\newblock A stochastic approximation method.
\newblock {\em The Annals of Mathematical Statistics}, 22:400--407, 1951.

\bibitem{sato2020}
Yuzuru Sato, Micka{\"e}l~D Chekroun, and Michael Ghil.
\newblock Convergence rate of snapshot attractors to random strange attractors.
\newblock {\em submitted}, 2020.

\bibitem{sato2018dynamical}
Yuzuru Sato, Thai~Son Doan, Jeroen~SW Lamb, and Martin Rasmussen.
\newblock Dynamical characterization of stochastic bifurcations in a random
  logistic map.
\newblock {\em arXiv preprint arXiv:1811.03994}, 2018.

\bibitem{sato2018analytical}
Yuzuru Sato, TS~Doan, NT~The, and HT~Tuan.
\newblock An analytical proof for synchronization of stochastic phase
  oscillator.
\newblock {\em arXiv preprint arXiv:1801.02761}, 2018.

\bibitem{skufca2006edge}
Joseph~D Skufca, James~A Yorke, and Bruno Eckhardt.
\newblock Edge of chaos in a parallel shear flow.
\newblock {\em Physical review letters}, 96(17):174101, 2006.

\bibitem{teramae2004robustness}
Junnosuke Teramae and Dan Tanaka.
\newblock Robustness of the noise-induced phase synchronization in a general
  class of limit cycle oscillators.
\newblock {\em Physical review letters}, 93(20):204103, 2004.

\bibitem{tsutsui2020center}
Daiji Tsutsui.
\newblock Center manifold analysis of plateau phenomena caused by degeneration
  of three-layer perceptron.
\newblock {\em Neural Computation}, 32:683--710, 2020.

\bibitem{wei2008dynamics}
Haikun Wei, Jun Zhang, Florent Cousseau, Tomoko Ozeki, and Shun-ichi Amari.
\newblock Dynamics of learning near singularities in layered networks.
\newblock {\em Neural Computation}, 20:813--843, 2008.

\bibitem{welling2011bayesian}
Max Welling and Yee~Whye Teh.
\newblock Bayesian learning via stochastic gradient langevin dynamics.
\newblock {\em In ICML}, pages 681--688, 2011.

\bibitem{xing2019walk}
Chen Xing, Devansh Arpit, Christos Tsirigotis, and Yoshua Bengio.
\newblock A walk with sgd: How sgd explores regions of deep network loss?
\newblock {\em In ICLR}, 2019.

\bibitem{zhu2019anisotropic}
Zhanxing Zhu, Jingfeng Wu, Bing Yu, Lei Wu, and Ma~Jinwen.
\newblock The anisotropic noise in stochastic gradient descent: Its behavior of
  escaping from sharp minima and regularization effects.
\newblock {\em In ICML}, 2019.

\end{thebibliography}

\end{document}